# International Comparative Studies on the Software Testing Profession

Luiz Fernando Capretz
*Dept Electrical & Computer Eng.*
*Western University*
London, Canada
lcapretz@uwo.ca

Pradeep Waychal
*Chhatrapati Shabu Institute of*
*Business Education and Research*
Kolhapur, India
pradeep.waychal@gmail.com

Jingdong Jia
*School of Software*
*Beihang University*
Beijing, China
jiajingdong@buaa.edu.cn

Yadira Lizama    Daniel Varona
*Cultureplex Laboratories*
*Western University*
London, Canada
{ylizama,dvaronac}@uwo.ca

*Abstract*—**This work attempts to fill a gap by exploring the human dimension in particular, by trying to understand the motivation of software professionals for taking up and sustaining their careers as software testers. Towards that goal, four surveys were conducted in four countries - India, Canada, Cuba, and China - to try to understand how professional software engineers perceive and value work-related factors that could influence their motivation to start or move into software testing careers. From our sample of 220 software professionals, we observed that very few were keen to take up testing careers. Some aspects of software testing, such as the potential for learning opportunities and the importance of the job, appear to be common motivators across the four countries, whereas the treatment of testers as second-class citizens and the complexity of the job appeared to be the most prominent de-motivators.**

*Keywords—software testing, testing professionals, testers, verification & validation.*

## I. INTRODUCTION

As software systems are becoming more ubiquitous, they are also becoming more susceptible to failures, resulting in potentially lethal combinations. Software testing is critical to preventing software failures, but is, arguably, the least understood part of the software life cycle and the toughest aspect of software development to perform correctly [1]. Adequate research has been carried out on both the process and technology dimensions of testing [2] [3], but not much on the human dimensions [4] [5].

One of the key components that has an impact on the performance and productivity of individuals in a job is the motivation to start and sustain it [6]. However, the field of software engineering—particularly software testing—is still in need of studies on motivation, especially the motivation to take up testing careers. Therefore, it is important to focus on phases of the software process, since there are considerable differences in the mindset and skills needed to perform different software tasks [7]. Regarding the need to study each phase of the software development process, Kanij et al. [8] and Ekwoge et al. [9] have discussed the lack of data available on addressing human factors in software testing. They observed that the current research on this topic has focused mainly on the development of testing methodologies and tools, but rarely on the human factors affecting professional software testers. Deak et al. [10] and Santos et al. [11] discussed the difference of opinions among software testers, regarding testing-related factors that could impact their motivation.

Software testers have different sets of roles and responsibilities. They should have a good understanding of the software system and its domain, which includes technical as well as product functionality. In order to create test cases it is also important that the software tester is aware of various testing techniques and which approach is best for a particular system. Finally, they should know what the various phases of the software life cycle are and how testing should be carried out in each phase. The responsibilities of the software tester include, but are not limited to:

- Analyzing the requirements from the client.
- Participating in preparing test plans.
- Preparing test scenarios, test cases and test data.
- Suggesting and preparing documents to improve the quality of the application.
- Communicating with the Test Lead / Test Manager.
- Participating in reviews and walkthroughs of testing procedures.
- Creating reports related to software testing that has been conducted.
- Documenting and tracking defects.
- Ensuring that all tested-related work is carried out as per the defined standards and procedures.

Ahmed et al. [12] analyzed hundreds of job advertisements for software engineers in many different countries all over the world. More than 50% of these ads asked for soft skills. Cerioli at al. [13] also analysed hundreds of job advertisement focusing on software testing and reported the technical skills desired for software testers. These studies show that companies want to hire testers who can communicate well and have technical as well as soft skills [14] [15]. They found that there is a significant increase in the need for openness, adaptability, the ability to work independently, and the ability to work as a team player. Additionally, new categories of soft skills are emerging, such as a strong work ethic, a customer-focussed mindset, and the ability to work under pressure.

While software engineering is delivering an unprecedented performance-to-cost ratio, it is also facing tough questions about glaring failures that have caused losses of billions of dollars and human lives [16]. However, very few bright individuals, across the globe, voluntarily choose testing careers [17] [18], which deprives the software industry of good testers, consequently leading to the delivery of poor quality products. To change this situation, it is necessary to analyze the reasons for such apathy towards testing careers.

## II. METHODOLOGY

Our study gathered and analyzed the opinions of software engineers about testing careers by asking a sample of professionals if they would choose testing careers and what they felt were the PROs and CONs of testing careers. Towards that end, we developed a cross-sectional but simple survey-based instrument, the questions are listed below:

1. What are the three Motivators/PROs (in the order of importance) for taking up a testing career?
a) _____________
b) _____________
c) _____________

2. What are three De-motivators/CONs (in the order of importance) for taking up a testing career?
a) _____________
b) _____________
c) _____________

3. What are chances of my taking up a testing career?
"Certainly Not" "No" "Maybe" "Yes" "Certainly Yes"
Reasons: _____________

The instrument was designed to collect responses on the motivation of software professionals for working as software testers, and to understand work-related factors in the specific context of software testing. Specifically, we asked professionals for the probability that they would choose testing careers by offering multiple choices: "Certainly Yes," "Yes," "Maybe," "No," and "Certainly Not". Since there has been limited prior research in the area, we decided to use a qualitative approach to investigate and understand the phenomena within their real-life context, and asked the respondents to provide an open-ended but prioritized list of PROs and CONs, and open-ended rationale regarding their decision to take up or not take up a testing career.

Our sample sought the views of 220 software professionals from four different countries on the PROs and CONs of testing careers. The choice to participate in the survey was voluntary. The Indian responses were collected from testing professionals working for various India-based companies, who had at least one year of experience in the profession. The Canadian responses were sought from alumni of a software engineering program at Western University with 1 to 10 years of experience. The Chinese subjects were software practitioners from different companies with 3 to 5 years of experience, studying part-time and pursuing their master's course at Beihang University in Beijing. The Cuban responses came from software professionals taking part-time graduate courses at the University of Informatics Science in Havana, they worked at research institutes, such as Centre for e-Government, affiliated to the university, and had 1-10 years of experience working in national and international projects. We, thus, used convenience sampling in terms of countries as well as software professionals.

Due to a varying number of respondents in the four geographies, we use percentage instead of absolute number of responses, increasing the validity of comparisons. We have included the PROs and CONs that were provided by at least 5% of professionals from each of the respective countries. Since we excluded PROs that were chosen by less than 5% of professionals, the total of each column may not be 100%.

## III. FINDINGS

This section presents the likelihood of software engineering professionals taking up testing careers and lists the PROs and CONs of all the respondents. The results of the third question on the likelihood of software engineering professionals taking up testing careers, are depicted in Table I. Our sample consisted of 220 software professionals from four different countries (22 from India, 20 from Canada, 34 from China, and 144 from Cuba). Since we were not studying the gender influence on the testing career decision, the response was option. Regarding women participation in the study, less than 10% of the Canadian subjects were female, about 15% of the Chinese professionals were female, and, surprisingly, 39% among Cubans; we could not figure out the percentage among Indian repondents because the survey was anonymous.

TABLE I.  CHANCES OF SOFTWARE ENGINEERING PROFESSIONALS TAKING UP TESTING CAREERS

| Response | Canada | China | Cuba | India |
|---|---|---|---|---|
| Certainly Not | 15% | 3% | 17% | n.a. |
| No | 30% | 23% | 47% | n.a. |
| Maybe | 30% | 59% | 15% | n.a. |
| Yes | 10% | 12% | 16% | n.a. |
| Certainly Yes | 15% | 3% | 6% | n.a. |

**PROs of testing careers as perceived by professionals**

The analysis of responses to the PROs resulted in the following categories.

- Learning opportunities
- Important job
- Easy job
- Thinking, creative, challenging, and interesting job

- More available jobs / Secure jobs / Stable jobs
- More monetary benefits
- Suitable for inducting freshers (new hires)
- Proximity to customers
- Increases commitment to product quality
- Good infrastructure

The responses from each country were analyzed and presented in Table II below. In all of the four geographic regions surveyed, we found that testing was not a popular career option among software professionals. Canada has the highest percentage of professionals (25%) who wanted to take up testing careers. Testing offers tremendous learning opportunities, as reported by professionals across the four countries. Barring Indian professionals, whose most voted PRO for testing was the thinking nature of the job, professionals from the other three countries voted learning opportunities as the most common PRO. Indians professionals voted that as the second PRO. The Chinese professionals' second PRO, on the other hand, was easiness of the job. With the exception of Cubans, other professionals also viewed the importance of testing jobs as another PRO. The Cuban PROs were barring learning opportunities, different, and included the suitability of testing jobs for inducting freshers, proximity to customers, and an increase in commitment to software quality. Further investigation of the reasons for such differences from the Cuban contingent is needed.

TABLE II. PERCENTAGE OF MOTIVATION DRIVERS FOR SOFTWARE TESTING PROFESSIONAS

| Motivators | Canada | China | Cuba | India |
|---|---|---|---|---|
| Learning opportunities | 34% | 36% | 45% | 28% |
| Important job | 16% | 7% |  | 28% |
| Easy job | 16% | 32% |  |  |
| Thinking job | 7% |  |  | 37% |
| More available jobs | 13% | 14% |  |  |
| More monetary benefits | 5% | 5% |  |  |
| Suitable for inducting freshers |  |  | 16% |  |
| Proximity to customers |  |  | 16% |  |
| Increases commitment to product quality |  |  | 13% |  |
| Good infrastructure |  |  | 5% |  |

**CONs of testing careers as perceived by professionals**

The analysis of responses to the CONs resulted in the following categories:

- Second-class citizen
- Lack of career progression
- Complexity / stressful / frustrating
- Tedious, less creative, not challenging
- Missed development / no coding
- Less monetary benefits
- Finding the mistakes of others
- Detail oriented skills

The survey respondents from each country were analyzed and are displayed in Table III below. The most common de-motivators (CONs) appeared to be the second-class citizen treatment meted out to the testers, along with complexity, resulting in stress and frustration. Except for the Indian professionals, others have concerns about career development and monetary benefits in testing tracks, and. with the exception of Cuban professionals, others were concerned about tediousness and missing development aspects of testing careers. Cuban professionals had different views and pointed out the difficulties of finding the mistakes of others and the requirement for detail oriented skills as CONs of a testing career. In fact, the "finding mistakes" reason was the most voted CON by the Cuban professionals.

TABLE III. PERCENTAGE OF DE-MOTIVATION DRIVERS FOR SOFTWARE TESTING PROFESSIONALS

| De-motivators | Canada | China | Cuba | India |
|---|---|---|---|---|
| Second-class citizen | 24% | 7% | 15% | 46% |
| Career progression | 22% | 15% | 7% |  |
| Complexity | 10% | 27% | 20% | 40% |
| Tedious | 17% | 25% |  | 6% |
| Missed development | 12% | 9% |  | 6% |
| Less monetary benefits | 10% | 9% | 13% |  |
| Finding mistakes of others |  |  | 23% |  |
| Detail oriented skills |  |  | 17% |  |

IV. DISCUSSIONS

This comparative study offers useful insights that can help global software industry leaders to come up with an action plan to put the software testing profession under a new light. That, in turn, could increase the number of software engineers choosing testing careers, which would promote quality testing. As discussed before, software testing is a human-dependent activity and the motivation of software testers can directly influence the quality of the final product.

In Table IV (motivators) and Table V (de-motivators) we compared our results across four different countries (Canada, China, Cuba, and India) with those listed by Deak et al. [10] in Norway and Santos et al. [11] in Brazil. The three studies used different words for motivators and de-motivators, but we have mapped them appropriately. Other studies have considered only students as subjects [19], thus not compared with experienced professionals in this study.

TABLE IV. COMPARING MOTIVATORS WITH OTHER STUDIES

| Our Study | Deak [10] | Santos [11] |
|---|---|---|
| Learning opportunities | | Acquisition of useful knowledge |
| Important jobs | Focus on quality improvement | |
| Easy jobs, work-life balance | | Well-defined work |
| More available jobs, secure jobs | | |
| Thinking, creative, challenging, and interesting job | Enjoy challenges, technically challenging work | Creativity |
| More monetary benefits | | |
| Suitable for inducting freshers | | |
| Proximity to customers | | |
| Increases commitment to product quality | | |
| Good infrastructure | | |
| | | Variety of work |
| | | Recognition |
| | | Good management |

TABLE V. COMPARING DE-MOTIVATORS WITH PRIOR STUDIES

| Our Study | Deak [10] | Santos [11] |
|---|---|---|
| Second-class citizen / less monetary rewards | Unhappy with management, lack of influence and recognition, time pressure | Recognition |
| Complexity / stressful / Frustrating / Requires more patience | Technical issues, time pressure | |
| Miss development / no coding | | |
| Tedious, less creative, not challenging | Boredom | Work variety |
| Career progression | | |
| More overtime | Unhappy with Management, lack of organization, time pressure | |
| Finding mistakes of others | Poor relationships with developers | |
| Requires detail oriented skills | | |
| Less monetary benefits | | |
| | | Working environment issues |

Our discovery of motivators and de-motivators for software testers can help global testing managers and team leaders who are dealing with motivational problems in prospective and current professionals in software testing. Further, these discussions can guide industrial practice in handling testing personnel issues that could affect software quality. By understanding the motivational and de-motivational factors, practicing managers may be able to attract larger numbers of professionals to testing careers as well as retain the current testing professionals.

V. CONCLUSIONS

The general empirical findings on the motivation to take up and continue with testing careers suggest that learning opportunities and importance of jobs appear to be common motivators across the four countries. The treatment of testing professionals as second-class citizens and the complexities that result in stressful and frustrating situations, appear to be common de-motivators. These findings match reasonably well with earlier studies that deal with motivation in testing jobs. Since we are studying the chances of taking up testing careers, monetary rewards and job availability have also appeared in the list.

Essentially the results of a study like this should be carefully used in different settings. We recommend that managers organize meetings with employees to help them to understand motivational and de-motivational factors specific to their situations and use the factors discovered by ours and other similar studies as a starting or reference point.

The software testing profession has been changing with the advent of Agile methods, DevOps and other paradigms. For instance, Developer in Test is a new role in many companies and this requires competencies other than that of a traditional tester from decades ago. Similarly, test automation, security testing, etc., are different forms of this profession. These aspects should be taken into account in future studies.

**Bio-Sketches:**

**Luiz Fernando Capretz** is a professor of software engineering at Western University in Canada and visiting professor of computer science at New York University in Abu Dhabi (UAE). His current research interests are software engineering, software testing, and human aspects of software engineering. Dr. Capretz received his Ph.D. from the University of Newcastle upon Tyne (U.K.), M.Sc. from the National Institute for Space Research (INPE-Brazil), and B.Sc. from UNICAMP (Brazil). He is a senior member of IEEE, a voting member of the ACM, a MBTI Qualified Practitioner, and a Certified Professional Engineer in Canada (P.Eng.). He can be reached at lcapretz@uwo.ca, and further information about him can be found at: http://www.eng.uwo.ca/people/lcapretz/ .

**Pradeep Waychal** is the director at CSIBER (www.siberindia.edu.in) and chairs Guruji Education Foundation (www.gurujifoundation.in) that provides holistic education to the underprivileged students. His research interest are at the intersection of human sciences and software engineering, engineering education, and innovation management. He has done PhD in developing innovation competencies for Information System Businesses from IIT

Bombay, India. He is a senior member of IEEE, a member of ACM and ASEE, and a life member of CSI (Computer Society of India). He was associated with organizations such as ISO-JTC1 SC7, NASSCOM, and ISBSG. Contact him at: pradeep.waychal@gmail.com.

**Jingdong Jia** is an associate professor of School of Software at Beihang University in China. She received her Masters and Ph.D. in management science and engineering from Beihang University. She has many years of experiences in software engineering practice and education. She is a member of CCF (China Computer Federation). Her research interests include requirements engineering, empirical software engineering, software testing and software engineering education. Contact her at: jiajingdong@buaa.edu.cn.

**Daniel Varona** is a Ph.D. student working at Cultureplex, in the Faculty of Arts and Humanities at Western University. He graduated in Software Engineering at the University of Informatics Sciences (UCI) in Havana, Cuba, where he taught computer science for eight years. His current research interests are in human factors on software engineering, empirical studies and HCI; and more recently he is working on machine learning´s limitations avoiding automation of bias. Contact him at: dvaronac@uwo.ca.

**Yadira Lizama-Mue** is a Ph.D. student of Hispanic Studies at the Faculty of Arts and Humanities, Western University and a research assistant and project manager at Cultureplex Lab (https://cultureplex.ca) where develops several projects and partnerships with companies in the London/Ontario area. She graduated in software engineering at the University of Informatics Sciences (UCI) in Havana, Cuba. Her research interests focus primarily on software development, network analysis, data science, and artificial intelligence to solve practical issues on the transitional justice and digital humanities fields. Contact her at: yadiralizama@gmail.com.